\documentclass[sigconf,dvipsnames,nonacm]{acmart}
\newif\ifanonymous
\anonymoustrue

\usepackage[noabbrev,nameinlink,capitalise]{cleveref}
\usepackage[inline,shortlabels]{enumitem}
\usepackage{dsfont}

\usepackage[separate-uncertainty=true,binary-units]{siunitx}
\newcommand{\reduce}{\textit{Reduce}}
\newcommand{\map}{\textit{Map}}

\newcommand{\percent}[1]{\SI{#1}{\percent}}
\newcommand{\frage}[1]{{{\color{blue}[\sf#1]}}}
\newcommand{\pesa}[1]{\frage{PS:#1}}
\newcommand{\luhu}[1]{\frage{LH:{\color{orange}#1}}}

\renewcommand{\frage}[1]{}

\AtBeginDocument{%
  \providecommand\BibTeX{{%
    \normalfont B\kern-0.5em{\scshape i\kern-0.25em b}\kern-0.8em\TeX}}}

\newif\ifcmyk
\cmykfalse

\input{setup/pgfplots.tex}
\usepackage{xspace}

\makeatletter
\newcommand{\whp}{w.h.p\@ifnextchar.{\@}{.\@\xspace}}
\newcommand{\iid}{i.i.d\@ifnextchar.{\@}{.\@\xspace}}
\makeatother

\usepackage{marvosym}

\newcommand{\restore}{ReStore}

\newcommand{\bigoh}{\mathcal{O}}

\begin{document}

\title{Scalable Fault-Tolerant MapReduce}

\newcommand{\affiliationkit}{
    \affiliation{
        \institution{Karlsruhe Institute of Technology (KIT)}
        \streetaddress{Kaiserstraße 12}
        \city{Karlsruhe}
        \country{Germany}
        \postcode{76133}
    }
}

\newcommand{\affiliationhits}{
    \affiliation{%
        \institution{Heidelberg Institute for Theoretical Studies (HITS)}
        \streetaddress{Schloss-Wolfsbrunnenweg 35}
        \city{Heidelberg}
        \country{Germany}
        \postcode{69188}
    }
}

\author{Demian Hespe}
\authornotemark[1] %
\orcid{0000-0001-6232-2951}
\affiliationkit

\author{Lukas Hübner}
\authornote{Authors contributed equally to this research.}
\email{huebner@kit.edu}
\orcid{0000-0001-9213-7597}
\affiliationkit
\affiliationhits

\author{Charel Mercatoris}
\authornotemark[1] %
\affiliationkit

\author{Peter Sanders}
\authornotemark[1] %
\orcid{0000-0003-3330-9349}
\affiliationkit

\begin{abstract}
    Supercomputers getting ever larger and energy-efficient is at odds with the reliability of the used hardware.
    Thus, the time intervals between component failures are decreasing.
    Contrarily, the latencies for individual operations of coarse-grained big-data tools grow with the number of processors.
    To overcome the resulting scalability limit, we need to go beyond the current practice of interoperation checkpointing.
    We give first results on how to achieve this for the popular MapReduce framework where huge multisets are processed by user-defined mapping and reducing functions.
    We observe that the full state of a MapReduce algorithm is described by its network communication.
    We present a low-overhead technique with no additional work during fault-free execution and the negligible expected relative communication overhead of $1/(p-1)$ on $p$ PEs\@.
    Recovery takes approximately the time of processing $1/p$ of the data on the surviving PEs.
    We achieve this by backing up self-messages and locally storing all messages sent through the network on the sending and receiving PEs until the next round of global communication.
    A prototypical implementation already indicates low overhead $<4\,\%$ during fault-free execution.
\end{abstract}



\maketitle

\section{Introduction and Related Work}\label{sec:intro}
Big Data processing frameworks like MapReduce~\cite{mapreduce2008} allow easy parallelization of complex computations on huge data sets.
A single MapReduce step processes data using two functions: \map{} and \reduce{}.
\map{} locally applies a user-defined function to every input item and outputs key-value-pairs.
\reduce{} gathers all items with the same key and applies a user-defined reduction function to all items with the same key.
Chaining multiple MapReduce steps with different \map{} and \reduce{} functions enables a wide range of computations.
This approach is appealing as users have to specify only these two functions while the framework takes care of parallelization, load balancing, and even fault-tolerance.
We define a failure as one or multiple PEs suddenly stopping to contribute to the computation (\emph{fail-stop}).
After a failure, the application has to redistribute its work among the remaining PEs (\emph{shrinking} recovery).

To the best of our knowledge, existing MapReduce implementations~\cite{mapreduce2008,shvachko2010hadoop,Condie2010MROnline,Ekanayake2010Twister,Guo2015FTMRMPI} create full checkpoint at each MapReduce step.\footnote{Current implementations write the checkpoints to a fault-tolerant distributed file system, but they could in priciple also be kept in memory.}
Besides incurring a considerable constant factor overhead during fault-free operation, this implies a fundamental limit to the scalability of MapReduce computations:
Consider machines with $p$ processing elements (PEs).
While the time to execute and checkpoint a MapReduce step grows with $p$, the time intervals between PE failures shrink.\footnote{Some systems already average over 2 failures/day~\cite{Gamell2014} and we expect future systems to fail every hour~\cite{Dongarra2015}.}
Eventually, errors are bound to occur in almost every MapReduce step and we would like to have a more fine-grained fault-tolerance mechanism which is able to tolerate such failures without a significant overhead.
In this brief announcement we present such a solution for the important case of single PE-failures in the fail-stop model using shrinking recovery.
\footnote{The approach is easy to generalize to groups of PEs (e.g., computers sharing the same power-supply).
    There are also at least probabilistic generalizations to multiple failures but one should keep in mind that creating checkpoints every few MapReduce steps might be a useful part of an overall system.
    See online supplement for details.}

\section{Method}\label{sec:method}

\begin{figure}
    \centering
    \includegraphics[width=\linewidth]{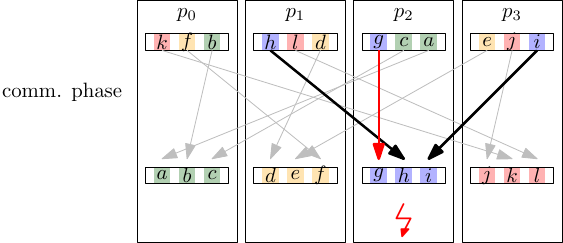}
    \caption{%
        Data flow and stored messages for MapReduce.
        Colors indicate destinations of data elements.
        Arrows indicate messages sent during the communication phase of a MapReduce step.
        After failure of PE $p_2$, we can reconstruct the data held by $p_2$ using the messages sent by other PEs.
        Black arrows correspond to messages available on the sending PE\@.
        The red self-message is the only message not available on another PE and has to be additionally communicated for backup.
    }\label{fig:self-messages}
    \Description[]{Data flow and stored messages for MapReduce}
\end{figure}

In the BSP model~\cite{valiant1990bsp} computations are performed in so called supersteps, where a local computation phase is followed by a synchronized message exchange phase.
Sanders~\cite{sanders2020connecting} describes how to implement MapReduce in the BSP~\cite{valiant1990bsp} model using two supersteps:
\begin{enumerate*}
    \item \map{}: Each PE maps its local elements and distributes the resulting key-value-pairs uniformly at random to all PEs (\textit{shuffle})
    \item \reduce{}: Each PE assembles the received elements to obtain the output set.
\end{enumerate*}
Note, that in the MapReduce model the data sent over the network during the shuffle at the end of the \map{} superstep fully describes the state of the program, which we can thus back up by storing those messages.
If we detect a failure during the subsequent MapReduce step, we can recover by re-executing the \map{} and \reduce{} operations on these stored messages (\cref{fig:self-messages}).

In order to use the communication phase of each MapReduce step as a recovery point, we
\begin{enumerate*}[label=(\alph*)]
    \item store all messages sent on the sending PE and
    \item send the self-messages of each PE $p_i$ to a different PE $b(i)$ for backup.
\end{enumerate*}
Upon failure of PE $p_i$, all other PEs send the data that they originally sent to $p_i$ to $b(i)$, so that $b(i)$ has all the data that $p_i$ received during the shuffle.
PE $b(i)$ can then recompute $p_i$'s result by applying the \reduce{} and \map{} functions locally (\cref{fig:self-messages}).

When using a single PE $b(i)$ as backup for each PE $p_i$, $b(i)$ holds twice as much data as all the other PEs after $p_i$ failed.
To avoid this imbalance, we divide each self-message into $p-1$ parts and send each part to a different PE.
This enables us to perform the recomputation in parallel on all remaining PEs after a failure and helps with load balancing of the next \map{} operation when continuing normal computation on $p-1$ processors.

In the BSP-based implementation~\cite{sanders2020connecting}, we choose the data destinations during the shuffle by hashing each key and splitting up the range of hash values evenly among the PEs.
During recovery, we further split up the range formerly assigned to the failed PE evenly among the surviving PEs and redistribute all messages sent to the failed PE (including the backed up self-message) accordingly.
We then merge the newly received items into the existing data after applying the reduction function and the subsequent \map{} operation.
At this point, the system is in a state from which it can continue normal operation on the remaining PEs.
As we require only the messages of the last shuffle for recovery, we can delete all previously stored messages after the subsequent \reduce{} operation.

\subsection{Analysis}\label{sec:analysis}

We analyze the running time of a single MapReduce step following the BSP-based analysis by Sanders~\cite{sanders2020connecting}.
Let $m$ be the total data volume and $\hat{m}$ be the maximum size of an input or output of the user-defined \map{} or \reduce{} functions.
For brevity, we omit the detailed discussion of total work $w$ and maximal work $\hat{w}$ that allow for analogous bounds.
\luhu{I re-added this, because we're referencing it in the abstract.}
The expected number of machine words across all self-messages is $m/p$.
Therefore, the expected overhead of the overall communication volume for enabling fault-tolerance, that is additionally copying self-messages over the network, amounts to $\frac{m}{m - m / p} - 1 = \frac{1}{p-1}$.
The worst imbalance in communication volume is obtained when all the data is concentrated on $m/\hat{m}$ objects of size $\hat{m}$ \cite{sanders2020connecting}.
We focus on a ``high-volume'' scenario where $m\in\Omega(\hat{m}p\log p)$, which implies that the bottleneck communication volume is $O(m/p)$ with high probability.
Moreover, the bottleneck communication volume with respect to single messages (e.g. the self-messages) will be
\begin{equation}\label{eq:message}
    \hat{m} \hat{o}(O(m/(\hat{m}p)), p)\approx O(m/p^2+\hat{m})
\end{equation}
where
$\hat{o}(b, p)$ is the expected maximum number of balls in a bin when uniformly at random assigning $\lceil b \rceil$ balls to $p$ bins. $O(m/p^2+\hat{m})$ is negligible compared to the bottleneck communication volume $O(m/p)$  of all other messages.\pesa{later: discuss: this analysis might be too conservative as the largest objects appearing in practice are on the input size of reducers which is not relevant for self-messages?}
Additionally, as backing up the self-messages is not on the critical path, we can overlap it with the computaitons of the subsequent \reduce{}.

\pesa{Comparing our MapReduce to the most widely used approach may be unfair as there is a lot of further research on the topic? Discuss.}
\luhu{I looked at the other approaches cited here and none of them seem to asymptotically improve the checkpointing overhead.}
In contrast, to the best of our knowledge, all existing fault-tolerant MapReduce implementations~\cite{mapreduce2008,shvachko2010hadoop,Condie2010MROnline,Ekanayake2010Twister,Guo2015FTMRMPI}
checkpoint the output of each \reduce{} phase, thus incurring $\bigoh(m)$ additional data to be sent over the network\footnote{In many cases, this data even ends up on a distributed file system, incurring further overheads -- in particular when the input of the next MapReduce step is also read from the file system.}.

During fault-free execution, neither we nor the reference MapReduce implementations process the additional data further, and do thus not incease the local work.

Neither our approach nor reference MapReduce create recovery points during the \map{} phases and thus have to re-execute it on the lost data after a failure.
In reference MapReduce, a failure of a PE during the \reduce{} causes the local work of the \textit{current} \map{} and \reduce{} of this PE to be re-executed.
Our method re-executes the local work of the \textit{preceeding} \reduce{} and \textit{current} \map{} (\Cref{sec:method}).
Similar to reference MapReduce, we distribute this recomputation across all surviving PEs (\Cref{sec:method}).
The failed PE received a total amount of $\bigoh(m/p)$ in the high-volume case.
As this is redistributed over the surviving $p-1$ PEs, no PE has to work on
more than \pesa{is this too much handwaving? What is really happening here?}$\bigoh(m/p^2+\hat{m})$ of this data (similar to Equation~(\ref{eq:message})).

In summary, enabling fault-tolerance occurs only a negligible communication and no local work overhead during normal operation and recovery takes approximately the time of processing a fraction $1/p$ of the data on $p$ PEs which is asymptotically negligible.

\section{Experiments}

\begin{figure}[t]
    \centering
    \scalebox{0.57}{
        \input{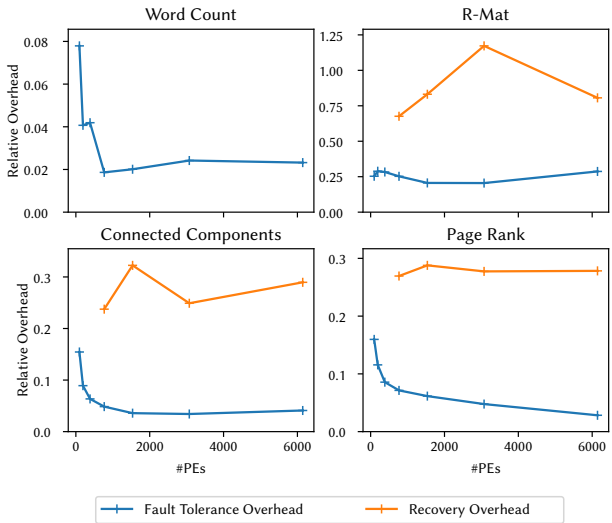}
    }
    \caption{Overhead for different MapReduce benchmark algorithms.
        ``Fault Tolerance'' shows the overhead caused by the backup mechanism in a fault-free scenario, i.e., relative to the running time without fault-tolerance enabled.
        ``Recovery'' shows the time taken to recover from a failure relative to the time of a single MapReduce step.
    }\label{fig:weak-scaling-overhead}
    \Description[]{Plots showing the runtime overhead when enabling fault-tolerance in our MapReduce implementation.}
\end{figure}

We evaluate the overhead caused by our fault-tolerance technique on four MapReduce benchmark algorithms: Word Count, R-Mat, Connected Components, and PageRank (\Cref{fig:weak-scaling-overhead}).
For details on the benchmarks algorithms and inputs, see the online supplement.

We use a prototype C++ implementation based on MPI with support for multi-PE-failures (see supplementary material for details).
In contrast to the theoretical description above, we send the self-messages of each PE to a single other PE instead of splitting it up among all remaining $p - 1$ PEs.
This does not change the bottleneck communication volume during normal operation but leads to some imbalance during recovery.
Further, we use the same hash function for every \reduce{} operation and do not distribute the outputs of the \reduce{} operation as used for the analysis.
This helps us to utilize locality in benchmark algorithms but can in turn lead to larger self-messages.

We run experiments on the SuperMUC-NG cluster as described in the online supplement.
We simulate failures of \percent{10} of the nodes used, always failing one node (48 PEs) at a time during the message exchange phase of the \reduce{} operation (at which real failures would be detected).
These failures are uniformly distributed across the MapReduce steps.
We show the average running time over five repetitions with different random seeds (\Cref{fig:weak-scaling-overhead}).

For low PE counts, a large fraction of the messages are self-messages, causing a large overhead for enabling fault-tolerance.
However, this is not a problem since fault-tolerance is of negligible importance for small systems as they fail less frequently.
Going to medium and high PE counts, the overhead reduces substantially.
For \num{6144} PEs, the overhead is \percent{2} for Word Count, \percent{3} for Page Rank, \percent{4} for Connected Components, and \percent{29} for R-MAT.
For all benchmarks except for R-MAT this is already reasonably low. The large overhead for R-MAT is due to the many self-messages caused by using the same hash function in every round.
Actual recovery is considerably more expensive (typically \percent{30} of the time of a normal MapReduce step) since our prototype performs the recovery on a single PE. Moreover, the \texttt{MPI\_Alltoallv} we use is know to scale poorly ~\cite{schlag2013distributed,Sanders2023Triangle,Sanders2023MST} and could be replaced by more efficient implementation.
Still, recovery is already much faster than repeating the complete MapReduce step.

\section{Conclusion and Future Work}\label{sec:conclusion}

We present a technique for low-overhead fault-tolerance and fast recovery in the general-purpose distributed MapReduce framework.
We empirically show that our technique causes low overhead for most benchmark algorithms and plan to add a comparison against other state-of-the-art implementations.
We plan to implement extensions for more efficient fault-tolerance: Creating recovery points less frequently could lower the overhead during fault-free execution, and splitting up self-messages during backup would improve load balance during recovery.
Many MapReduce implementations support additional features: For example, data which does not change across MapReduce steps does not need to be sent over the network.
We could recover this static data after a failure using \restore{}~\cite{RestoreConferenceVersion}; which we could also use for the zero-overhead variant explained in the online supplement.
Our MapReduce implementation itself could be extended by adding load balancing (e.g., randomized or work-stealing~\cite{sanders2020connecting}), local aggregation for reduction functions that allow it, improved all-to-all communication (e.g., 1-factor algorithm~\cite{Sanders2002OneFactor,schlag2013distributed} or grid-based ~\cite{Sanders2023MST,Sanders2023Triangle}), and a hybrid parallelization for better utilization of the multicore machines in modern HPC clusters.
The basic technique could also be generalized for the richer operation sets in tools like Spark \cite{Spark} or Thrill \cite{BAJLNNSSSS16}.

\begin{acks}
    We gratefully acknowledge the Gauss Centre for Supercomputing e.V\@. (\url{www.gauss-centre.eu}) for funding this project by providing computing time on the GCS Supercomputer SuperMUC-NG at Leibniz Supercomputing Centre (\url{www.lrz.de}).
\end{acks}

\bibliographystyle{ACM-Reference-Format}
\bibliography{references-filtered}

\end{document}


\title{Online Supplement to\\Scalable Fault-Tolerant MapReduce}

\maketitle

\section*{Lowering the Checkpoint Frequency}\label{method:lowering-cp-freq}

\begin{figure}[htb]
    \centering
    \includegraphics[width=0.7\linewidth]{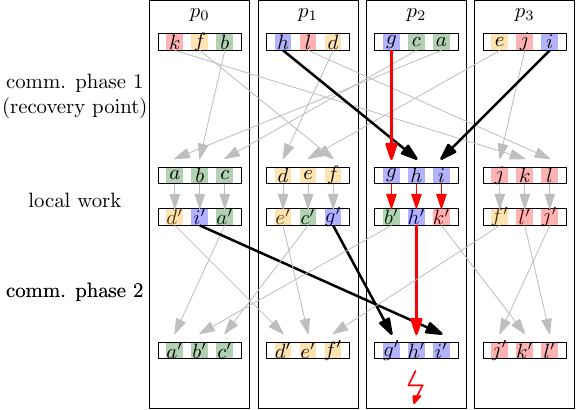}
    \caption{
        Messages lost if a shuffle phase is not used as a recovery point, i.e., we do not back up self-messages at other PEs.
        After failure of $p_2$ we need the data from all red arrows for recovery.
        Self-messages from communication phase 1 are backed up so we can recompute data elements $b', h'$, and $k'$ from the stored messages.
        These recomputed data elements in combination with the stored messages from the other PEs are used for recovery of communication phase 2.
        During recovery, we do not re-send data elements $b'$ and $k'$ because they were not sent to $p_2$~before.}\label{fig:recovery-points}
\end{figure}

In order to reduce the overhead for copying the self-messages over the network during every shuffle, we can change some shuffle phases to non-recovery-points by storing just the messages sent without backing up the self-messages.
During recovery we then need to recompute the data lost starting at the last recovery point.
We recover the last \reduce{} step that \emph{did} include a recovery point as before.
All following \reduce{} steps without a recovery point use the recovered data as during normal execution: On each PE, the data previously sent to the failed PE $F$ along with the recovered data is used to re-execute the operations.
During the communication phases we have to discard any parts of the recovered data that lies outside of $F$'s hash-range, i.e., not send it to the destination PE as it was already sent before the failure (\Cref{fig:recovery-points}).
This technique could be taken to the extreme by never backing up self-messages.
In that case, after a failure, the data originally obtained by $F$ from the data source (or a checkpoint created in between) would have to be re-read (usually from a fault-tolerant file system like HDFS) and all operations would have to be re-executed on the lost data.
This would cause virtually zero runtime overhead during fault-free execution but would substantially increase recovery times and the memory overhead for storing all messages sent to other PEs.
This would also enable recovering from any number of simultaneous PE failures because all data of all failed PEs can be recovered.

\section*{Supporting Multiple Failures}\label{sec:multinode}

Consider two PEs $p_i, p_j$ failing at the same time. Even if $p_i$ does not back up the self-messages of $p_j$ (and vice-versa), the messages sent from $p_i$ to $p_j$ are lost.
We can extend our technique to support failures of predefined groups of PEs, e.g., all CPUs on the same power supply, which are more likely to fail at once than unrelated PEs~\cite{BautistaGomez2011} and multiple single failures between recovery points.
We archive this by logically treating them as one PE and treating all messages sent between them as self-messages, we can apply the same technique as described in the main text.

\section*{Experimental Setup and Environment}\label{sec:experiments}

Our code is written in C++ and compiled using GCC 8.4.0 with full optimizations (\texttt{-O3}).
We ran all our experiments on the SuperMUC-NG super computer.\footnote{\url{https://doku.lrz.de/display/PUBLIC/SuperMUC-NG}}
Each node consists of two Intel Skylake Xeon Platinum 8174 processors with \(24\) cores and \(96\) GB of memory each, connected via an OmniPath network with a bandwidth of~\SI{100}{\giga\bit\per\second}\@.
The operating system is SUSE Linux Enterprise Server 15 SP1 running Linux Kernel version 4.12.14-197.78. %
Unless otherwise stated, we communicate using OpenMPI version 4.0.4.
The recent MPI 4 standard includes mechanisms for detecting failures and re-establishing a consistent environment after a failure. An implementation called ``ULFM'' is available for OpenMPI~\cite{Bland2013}.
We verify that our implementation functions properly if nodes actually fail and communication is recovered with ULFM as part of our fully automated unit tests.
The current version of ULFM at the time of running our experiments, however, was not stable enough to conduct reliable performance benchmark experiments.
For example, processes may be reported incorrectly as failed or recovery may result in two separate groups of nodes that each assume that the other group has failed.
We reported this behavior on the ULFM mailing list and the authors of ULFM reproduced and confirmed the bug.\footnote{George Bosilca. Post \texttt{pbSToy94RhI/xUrFBx\_1DAAJ} on the ULFM mailing list.}
Additionally, most communication and fault-tolerance mechanisms are currently slow (see Hübner \etal~\cite{Huebner2021} for details).
We expect these issues in ULFM to be fixed once more resources are allocated to implementing these features.
In our performance benchmarks, we thus use OpenMPI and simulate failures by removing processes from the calculation using \texttt{MPI\_Comm\_split} and replacing other required fault recovery steps by functionally similar ones (e.g., replacing \texttt{MPIX\_Comm\_agree} with \texttt{MPI\_Barrier}s).

\section*{Benchmark Algorithms}\label{sec:experimental-algorithms}

We evaluate the overhead caused by our fault-tolerance technique on four popular MapReduce benchmark algorithms: Word Count, R-Mat, Connected Components, and PageRank:

\textbf{Word Count}'s input is a text which is split up into words during the \textit{Map} phase.
In the \reduce{} phase Word Count gathers each occurrence of a word and counts the number of times this word occurs in the input text.
This is not an iterative algorithm, so there are no iterations to insert faults into but we still show it here due to its popularity.
As an input we use \SI{2}{\giga\byte} of text per node, generated by picking words uniformly at random from the $\num{4436574}$ distinct words in Project Gutenberg~\cite{ProjectGutenberg}.

\textbf{R-Mat}, or Recursive Matrix Model~\cite{r-mat2004}, is a random graph generator.
Given a number of vertices and edges, R-MAT generates each edge based on a probability distribution in the adjacency matrix.
It is commonly used to generate graphs with a community structure and is used in the Graph 500 benchmark~\cite{murphy2010introducing}.
We use the MapReduce implementation of the R-Mat generator introduced by Plimpton and Devine~\cite{PlimptonDevine2011MapReduce}.
The input is a set of edges produced by the probability distribution that may contain duplicate edges.
The \map{} phase has no functional role in this benchmark algorithm.
In the \reduce{} phase, we collect all edges with the same two endpoints and generate new ones for every duplicate.
We repeat these MapReduce steps until all edges are unique.
We generate graphs with $2^{18}$ vertices per node and an average degree of $30$ using the same randomization parameters used in the Graph 500 benchmark.

\textbf{Connected Components} are maximal connected subgraphs and a basic tool for graph analysis and a building block of many graph algorithms.
We implement an algorithm by Kiveris~\etal~\cite{Kiveris2014CCs} which consists of two phases that are repeated until convergence.
Both phases output edges in a specific vertex order (i.e., directing the edge from vertex $u$ to $v$ or from $v$ to $u$) during the \map{} phase.
We then group the edges by the first endpoint during the \reduce{} phase where they are redirected to their connected components representative.
We repeat these MapReduce steps until convergence.
We compute connected components of random graphs~\cite{erdHos1960evolution} with $2^{18}$ vertices per compute node and an average degree of $0.5$.
This leads to graphs with many small connected components~\cite{erdHos1960evolution}.

\textbf{PageRank} is a network analysis algorithm developed by Google~\cite{page1999pagerank} which simulates a random walk on a web graph.
Each vertex of a graph starts with the same score with the sum of scores adding up to $1$.
The algorithm then works in rounds where each vertex's current score is split up and transferred in equal parts to each neighbor.
Additionally, each vertex gets the same small amount of score every round.
These two components are always normalized to keep the global sum at $1$.
We implement this in MapReduce by outputting for each neighbor of a vertex, the score transferred to that neighbor in the \map{} phase.
In the \reduce{} phase, the scores sent to each vertex are gathered, summed and a constant offset is added to simulate random jumps.
In order to keep all required information across rounds, we additionally emit the neighborhood of each vertex in the \map{} phase.
This neighborhood is then also output during the \reduce{} phase together with the score.
We run the PageRank algorithm for $100$ iterations on random graphs~\cite{erdHos1960evolution} with $2^{18}$ vertices per node and an average degree of $38$.

\bibliographystyle{splncs04}
\bibliography{references-filtered}